\documentclass[a4paper]{article}
\usepackage{graphicx} 
\usepackage[left=2.54cm,right=2.54cm,top=1.905cm,bottom=2.2225cm]{geometry}

\usepackage[english]{babel}

\usepackage[T1]{fontenc}
\usepackage{lmodern}
\usepackage{color}
\usepackage{hyperref}

\usepackage{mathtools}
\usepackage{amssymb}
\usepackage{bm}
\usepackage{upgreek}
\usepackage{bbold}
\usepackage{physics}
\usepackage[only,llbracket,rrbracket]{stmaryrd}

\usepackage{cleveref}
\usepackage{csquotes}
\usepackage{float}
\usepackage{placeins}
\usepackage{adjustbox}
\usepackage{booktabs}
\usepackage{caption}

\usepackage{titlesec}
\titlespacing*{\paragraph}{0pc}{0.4ex plus .1ex minus .2ex}{0.5em}

\newcommand{\R}{\ensuremath{\mathbb{R}}}

\DeclareMathOperator*{\argmin}{arg\,min}
\DeclareMathOperator*{\minimize}{minimize}

\usepackage{xparse}

\usepackage[style=numeric-comp,url=true,sorting=none]{biblatex}
\renewbibmacro*{doi+eprint+url}{%
    \printfield{doi}%
    \newunit\newblock%
    \iftoggle{bbx:eprint}{%
        \usebibmacro{eprint}%
    }{}%
    \newunit\newblock%
    \iffieldundef{doi}{%
        \usebibmacro{url+urldate}}%
        {}%
    }
\usepackage{authblk}

\title{{\large Submission to the 2024 Airbus/BMW Quantum Computing Challenge:}\\ Quantum-assisted Stacking Sequence Retrieval and Laminated Composite Design}
\author[1]{Arne Wulff}
\author[1]{Swapan Madabhushi Venkata}
\author[1]{Boyang Chen}
\author[2]{Sebastian Feld}
\author[2]{Matthias M\"oller}
\author[1]{Yinglu Tang}
\affil[1]{\textit{Faculty of Aerospace Engineering, TU Delft}}
\affil[2]{\textit{Faculty of Electrical Engineering, Mathematics \& Computer Science, TU Delft}}
\date{}

\begin{document}

\nocite{Wulff2024}

\maketitle

\begin{abstract}
    We, the \textit{QAIMS}\footnote{\href{https://www.tudelft.nl/lr/qaims}{https://www.tudelft.nl/lr/qaims}} lab at the Aerospace Faculty of TU Delft, participated as finalists in the \textit{Airbus/BMW Quantum Computing Challenge 2024: The Quantum Mobility Quest}\footnote{\href{https://thequantuminsider.com/2024/06/12/finalists-named-for-the-airbus-bmw-group-quantum-computing-challenge/}{https://thequantuminsider.com/2024/06/12/finalists-named-for-the-airbus-bmw-group-quantum-computing-challenge/}} in the ``Golden App'' category. Stacking sequence retrieval, a complex combinatorial task within a bi-level optimization framework, is crucial for designing laminated composites that meet aerospace requirements for weight, strength, and stiffness. This document presents the scientifically relevant sections of our submission, which builds on our prior research on applying quantum computation to this challenging design problem \cite{Wulff2024}.

    For the competition, we expanded our previous work in several significant ways. First, we incorporated a full set of manufacturing constraints into our algorithmic framework, including those previously established theoretically but not yet demonstrated, thereby aligning our approach more closely with real-world manufacturing demands. We implemented the advanced Filtering Variational Quantum Eigensolver (F-VQE), which enhances the probability shaping of optimal solutions, improving on simpler variational quantum algorithms. Our approach also demonstrates flexibility by accommodating diverse objectives -- such as buckling resistance maximization and novel applications of nearest-neighbor interactions for ply-angle clustering and dispersion -- as well as finer ply-angle increments ($15^\circ$ separation) alongside the previously demonstrated conventional ply angles.
    
    Scalability was tested using the Density Matrix Renormalization Group (DMRG) algorithm, which, despite limitations in entanglement representation, enabled simulations with up to 200 plies. Results were directly compared to conventional stacking sequence retrieval algorithms -- LAYLA and Opti-BLESS -- with DMRG showing high competitiveness. Given DMRG's limited entanglement capabilities, it serves as a conservative baseline, suggesting potential for even greater performance on fully realized quantum systems.
    
    This document serves to make our competition results publicly available as we prepare a formal publication on these findings and their implications for aerospace materials design optimization.
\end{abstract}

\clearpage
\section{Submission Summary}

In this work, we propose the ``Golden Application'' of quantum computing for advanced lightweight aerospace and automotive structures -- the problem of stacking sequence retrieval (SSR), a key step in the optimization loop of fiber-reinforced composite laminates. We are the first team in the world to have tackled the SSR problem from a QC angle, the work on which has been recently published in journal \textit{Computer Methods in Applied Mechanics and Engineering} \cite{Wulff2024}. There, we have translated the constrained SSR problem into a quantum setting and performed numerical demonstrations using both Variational Quantum Algorithms and the Density-Matrix Renormalization Group algorithm. A summary of this work has been submitted in the first round of this QC Challenge. In this second round, building upon our previous work, we present here:
\begin{itemize}
    \item an introduction of the bi-level methodology for the optimization of composite laminates and the motivation for the ``Golden App'' -- SSR problem;
    \item our quantum formulation of the SSR problem including all the manufacturing constraints; 
    \item a novel clustering-dispersion constraint to consider variable-thickness ply-blocks;
    \item the original objective function in our earlier work and a new objective function for buckling factor maximization;
    \item two proposed solvers based on the Density-Matrix Renormalization Group (DMRG) algorithm and the Filtering Variational Quantum Eigensolver (F-VQE) algorithm, respectively;
    \item results of the DMRG and F-VQE solvers on a range of test cases;
    \item comparisons of DMRG against two popular classical methods on various aspects such as accuracy and computational time; 
    \item scalability of our DMRG algorithm in terms of number of plies and number of ply-angle choices; and
    \item discussions of results, conclusions, and future outlooks.
\end{itemize}
In the rest of the report, we will explain the above list of items in detail. In the end, we will demonstrate that our proposed methodology:
\begin{itemize}
    \item outperforms the classical methods in terms of accuracy and efficiency on most test cases;
    \item shows excellent scalability when increasing the number of plies and the number of ply-angle choices;
    \item is flexible in terms of including additional constraints such as those on buckling factor and vibration frequency; 
    \item can include variable-thickness ply-blocks, thereby establishing a direct connection to manufacturing costs associated with tape-laying (thicker ply-blocks allow faster tape-laying to reach the desired laminate thickness); and
    \item opens the door to many possibilities of future applications and extensions, ranging from in-situ composites failure analysis during optimization to the design of multi-functional materials. 
\end{itemize}

\clearpage
\section{Detailed Explanation}
\noindent\href{https://github.com/ArneWulff/quantum-mobility-quest-qaims}{\raisebox{-0.4em}{\includegraphics[height=1.5em]{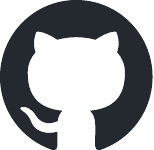}} \hspace{0.5em}\texttt{github.com/ArneWulff/quantum-mobility-quest-qaims}}

\subsection{Background and motivation}

As the mobility sector strives for carbon-neutrality, the weight of a vehicle remains one of the most critical factors affecting both its fuel consumption and overall environmental impact. To meet the varied demands of complex, multidirectional loadings, a material with customizable mechanical properties in each direction has the potential to reduce weight more effectively than isotropic alternatives, enabling efficient designs without sacrificing safety. For this reason, composites -- lightweight materials made of fibers embedded in a matrix -- are becoming the material of choice for weight reduction in the transportation industry. Their high specific strength in the fiber direction and possibility to layer them in various orientations, as laminated composites, allow for tailored stiffness and strength characteristics to meet the demands of multidirectional loadings. 

\begin{figure}
    \centering
    \includegraphics[scale=0.9]{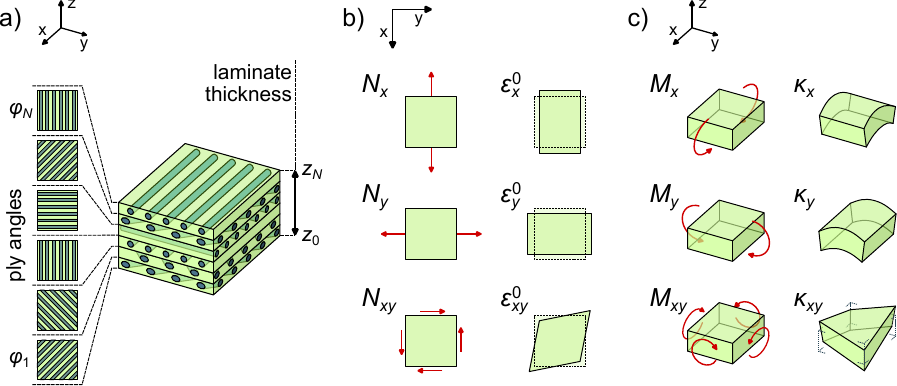}
    \caption{a) Diagram of a laminated composite material, which consists of multiple layers that are oriented in different directions. b) A representation of the stress resultants $\vec{N}$, which for isotropic materials result in in-plane deformations $\vec{\varepsilon}^{\: 0}$. c) A pictographical representation of the moment resultants $\vec{M}$, which for isotropic materials result in out-of-plane bending $\vec{\kappa}$. Non-diagonal elements in the $\mathbf{ABD}$-matrix allow for coupling between all possible stress and moment resultants on the one hand, and all in- and out-of-plane deformations on the other hand.}
    \label{fig:laminate}
\end{figure}

\begin{figure}
    \centering
    \includegraphics[scale=0.9]{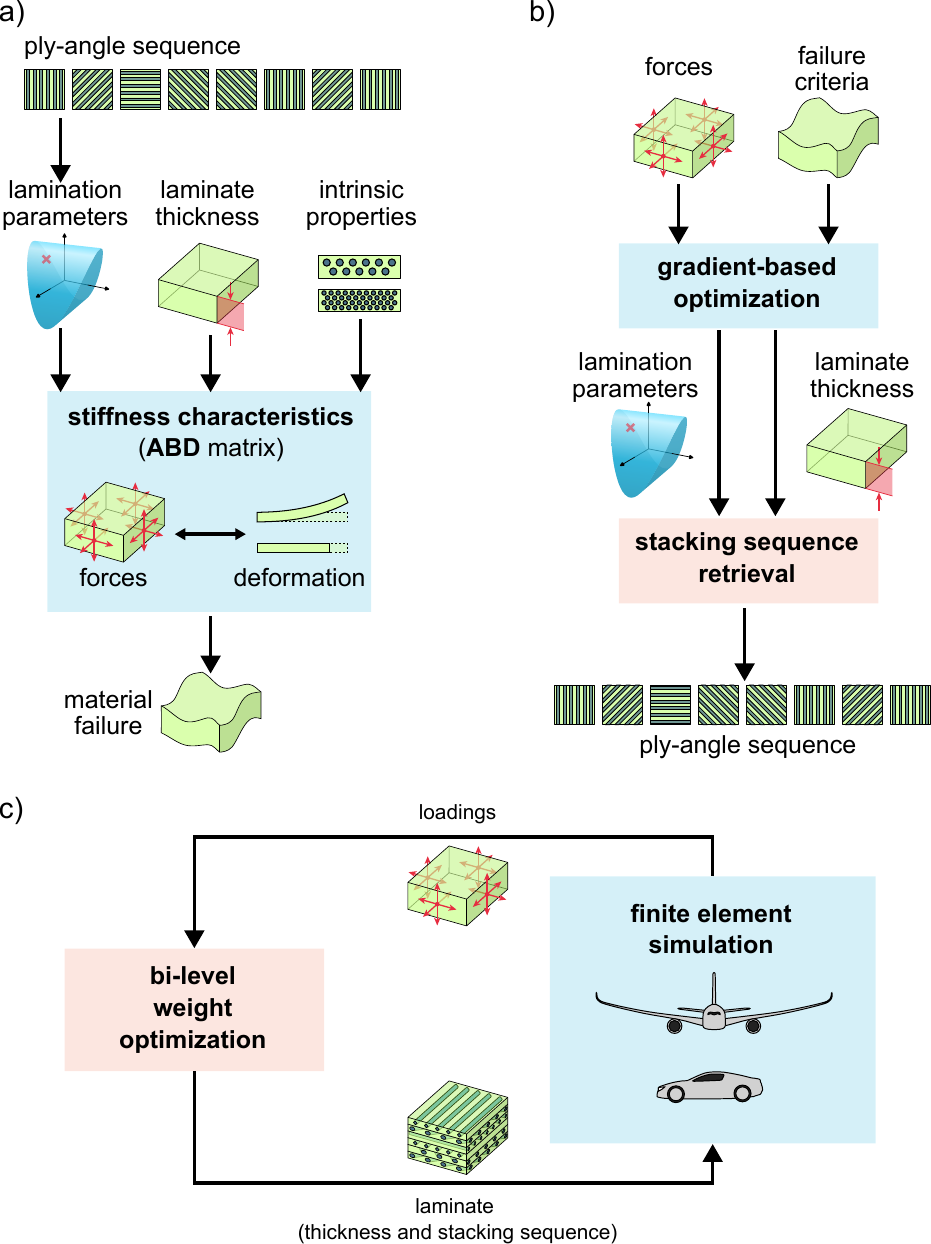}
    \caption{a) A depiction of the dependencies of the stiffness matrix which couples forces to deformations, and in turn is used to define failure criteria. b) A depiction of the bi-level optimization process for laminated composites. Additional manufacturing constraints that also may enter stacking sequence retrieval are not shown. c) Laminate design is often part of an outer structural design loop together with mechanical simulations.}
    \label{fig:bilevel}
\end{figure}

The classical lamination theory, based on the classical plate theory, is usually used to describe the behavior of the laminate. To a linear approximation, this behavior is commonly characterized by the stiffness or $\mathbf{ABD}$ matrix, which relates the stress and moment resultants $\vec{N}$ and $\vec{M}$ to the in-plane and out-of-plane deformations $\vec{\varepsilon}^{: 0}$ and $\vec{\kappa}$ \cite{Reissner1961,Dong1962,Tsai1980}:
\begin{equation} \label{eq:hook}
    \begin{bmatrix}
    \vec{N} \\ \vec{M}
\end{bmatrix} = \begin{bmatrix}
    \mathbf{A} & \mathbf{B} \\ \mathbf{B} & \mathbf{D}
\end{bmatrix} \begin{bmatrix}
    \vec{\varepsilon}^{\: 0} \\ \vec{\kappa}
\end{bmatrix}.
\end{equation}
The stiffness matrix is determined by the laminate thickness $h$, the intrinsic material properties of a single ply, and the ply angle sequence $\vec{\phi} = (\phi_1, \phi_2, \dots, \phi_N)$, i.e., the stacking sequence (see Figures \ref{fig:laminate} and \ref{fig:bilevel} a in \cref{SuppLaminate}).

However, this adaptability in laminated composites comes with a cost: each ply angle adds a degree of freedom to the design of the laminate. With potentially tens or hundreds of plies, optimizing these angles together with the laminate thickness to meet the stiffness and strength requirements, particularly when subjected to nonlinear failure criteria, becomes a highly challenging task \cite{Ghiasi2009,Ghiasi2010,Nikbakt2018,Nikbakht2019}. 

To reduce this complexity, one can leverage that different stacking sequences can yield the same $\mathbf{ABD}$ matrix, whose essential information is captured by at most 12 lamination parameters on which the matrix depends linearly:
\begin{align} \label{eq:abd_from_lp_intro}
\mathbf{A} &= h \qty(\bm{\Gamma}_0 + \bm{\Gamma}_1 v_1^A + \bm{\Gamma}_2 v_2^A + \bm{\Gamma}_3 v_3^A + \bm{\Gamma}_4 v_4^A),\\
\mathbf{B} &= \frac{h^2}{4} \qty(\bm{\Gamma}_0 + \bm{\Gamma}_1 v_1^B + \bm{\Gamma}_2 v_2^B + \bm{\Gamma}_3 v_3^B + \bm{\Gamma}_4 v_4^B), \notag\\
\mathbf{D} &= \frac{h^3}{12} \qty(\bm{\Gamma}_0 + \bm{\Gamma}_1 v_1^D + \bm{\Gamma}_2 v_2^D + \bm{\Gamma}_3 v_3^D + \bm{\Gamma}_4 v_4^D), \notag
\end{align}
where $v_l^X$ (with $l\in\{1,2,3,4\}, X\in\{A, B, D\}$) are the lamination parameters which encode the stacking sequence, and the matrices $\bm{\Gamma}_l$ are composed of the Tsai-Pagano material invariants independent of the stacking sequence \cite{Tsai1968,IJsselmuiden2011,Bordogna2020}. For the detailed expressions of these quantities, please refer to the supplementary \cref{SuppLaminate}.

These lamination parameters are continuous, as opposed to the ply angles which usually come with discrete choices due to practical manufacturing limitations. They also form a convex space. These amenable properties of the lamination parameters allow for a widely used bi-level optimization strategy \cite{Yamazaki1996,IJsselmuiden2009,Liu2013,Liu2019, Albazzan2019,Bordogna2020} (see \Cref{fig:bilevel} b and c in \cref{SuppLaminate}): the first step optimizes these lamination parameters and the overall laminate thickness to meet the stiffness and strength constraints, reducing the problem to a manageable size and enabling the use of continuous, convex optimization methods. The second step then finds a stacking sequence whose corresponding lamination parameters match the optimized ones obtained in step one as closely as possible while satisfying the manufacturing constraints. Ideally, this two-step design approach would yield a stacking sequence and thickness configuration that minimizes weight while satisfying all constraints.

While the bi-level optimization greatly simplifies the problem by focusing on lamination parameters, the second step, i.e., stacking sequence retrieval, remains a difficult task. As explained earlier, the ply angles are selected from a discrete set rather than freely chosen, which transforms the problem into a combinatorial optimization with an exponentially growing configuration space as the number of plies increases. Moreover, the manufacturing constraints impose additional rules for the retrieval procedure, such as limiting consecutive identical plies or preventing certain ply angles from being adjacent, which adds further to the complexity \cite{Niu1988,Bailie1997,compmathandbook2002,Albazzan2019}. As a result, it is generally hard to find a stacking sequence that closely matches the desired stiffness and weight requirements, which is crucial for achieving the full potential of weight reduction and performance. The combinatorial nature of this task, however, presents an opportunity for quantum computational methods, which hold promise for addressing these kinds of optimization problems in the future \cite{Lucas2014,Farhi2014,Farhi2016,Amaro2022}. In this work, we build upon our earlier work \cite{Wulff2024} and continue our exploration of quantum methods for stacking sequence retrieval. With its potential to significantly improve the design and optimization of laminated composites, this approach may very well represent a ``Golden App'' for quantum computing in materials and structural design, offering substantial benefits to vehicle weight reduction and carbon-neutral mobility.

\subsection{Problem description}
\subsubsection{Stacking sequences, lamination parameters and constraints}\label{subsec:constraints}

Following our previous publication \cite{Wulff2024}, we index the allowed discrete ply angles $\{\theta_1,\theta_2,\dots,\theta_d\}$ and represent the stacking sequence $\vec{\theta} = (\theta_{s_1}, \theta_{s_2}, \dots, \theta_{s_N})$ using indices $\vec{s} = (s_1,s_2,\dots,s_N)$, where $s_n \in \{1,\dots,d\}$. The lamination parameters $v_l^X$ share a common form:
\begin{equation} \label{eq:lp_gen_intro}
    v^X_l(s_1,\dots,s_N) = \sum_{n=1}^N \alpha_n^X f_l(s_n). 
\end{equation}

In this work, we focus on symmetric laminates, for which the $B$-lamination parameters vanish, reducing the number of relevant lamination parameters to 8. For most of our demonstrations, we selected the conventional ply angles ${0^\circ, +45^\circ, 90^\circ, -45^\circ}$ as the allowed ply angles for the laminate. In this case, the lamination parameters for $l=4$ are also zero, leaving us with only 6 lamination parameters to consider: $v_1^A, v_2^A, v_3^A, v_1^D, v_2^D, v_3^D$. For our comparative study, we additionally conducted experiments with a finer set of ply angles, spaced at $15^\circ$ intervals, respectively, where all 8 lamination parameters are required.

For a full set of manufacturing constraints, we used the following \cite{Albazzan2019,Fedon2021}:
\begin{itemize}
\item A disorientation constraint, allowing neighboring plies to differ in angle by at most $45^\circ$.
\item A contiguity constraint, limiting the number of consecutive plies with the same orientation to 5.
\item A balanced laminate condition, where for all angles except $0^\circ$ and $90^\circ$, the number of plies with angle $\phi$ must equal the number of plies with angle $-\phi$.
\item A 10\% rule, ensuring that the ply angles $0^\circ$, $90^\circ$, and $\pm 45^\circ$ each constitute at least 10\% of the total ply-angle sequence.
\end{itemize}

The contiguity constraint above is motivated from a damage-resistance point of view, as thinner ply-blocks manifest higher in-situ strengths and tend to suppress delamination. During the feedback sessions with the Airbus and BMW experts, the possibility of including variable-thickness ply-blocks in the optimization framework was brought up from a manufacturing point of view -- laying a laminate with thicker ply-blocks is more cost-effective, making it desirable to cluster same-angle plies into larger blocks. Therefore, we also explored how a nearest-neighbor coupling can control the clustering and dispersion of plies with the same angle, providing a probabilistic alternative to the contiguity constraint described above. This clustering or dispersion can be adjusted by the sign of the bias in the nearest-neighbor interaction. In this way, we allow designers to tune the relative importance of the different view-points with respect to ply-block thickness, allowing them to experiment with different trade-offs. Additional details can be found in the supplementary materials \cref{SuppCluster}.

\subsubsection{Optimization objectives}

We employed our approach for finding stacking sequences with the same optimization objective as in our previous work \cite{Wulff2024}: Given target lamination parameters $\xi_l^X$ and a specified ply number $N$, which are determined in the first step of the bi-level optimization, the goal of the second step -- stacking sequence retrieval -- is to find a stacking sequence $\vec{s}$ that satisfies the manufacturing constraints and matches the target lamination parameters as closely as possible:
\begin{equation}
\minimize_{\vec{s}, \ \vec{s} \ \text{valid}} \ \norm{\vec{v}(\vec{s}) - \vec{\xi}}.
\end{equation}
In this work, depending on the algorithm use, we employ either the Euclidean distance:
\begin{equation} \label{eq:euclid}
\mathrm{dist}_{\vec{\xi}}(\vec{s}) = \sqrt{\sum_{X,l} \qty(v_l^X(\vec{s}) - \xi^X_l)^2},
\end{equation}
or its square:
\begin{equation}\label{eq:euclid2}
\mathrm{dist}_{\vec{\xi}}^2 (\vec{s}) = \sum_{X,l} \qty(v_l^X(\vec{s}) - \xi^X_l)^2,
\end{equation}
to quantify the distance to the target lamination parameters.

Our approach can also be adapted to other objectives, such as buckling factor or fundamental frequency maximization (see \cref{eq:lambdaB} and \cref{eq:fundafreq} in \cref{SuppLaminate}), which were brought up in the feedback sessions. For conventional ply angles $\{0^\circ, +45^\circ, 90^\circ, -45^\circ\}$, the number of plies in the laminate with a specific angle can be calculated explicitly from the $A$-lamination parameters. We can then permute the positions of the plies along the thickness direction of the laminate \cite{Liu2000} to obtain the stacking sequence whose $D$-lamination parameters optimize the fundamental frequency or buckling factor. To demonstrate this novel adaptation, we maximized the buckling factor for bi-axial loading as an alternative objective with some of the algorithms we demonstrate in the later section. For more technical details, we refer to the supplementary materials \cref{SuppLaminate}.

\subsection{Methods}

\subsubsection{Representing stacking sequences and optimization objectives}

In order to represent the stacking sequence $\vec{s} = (s_1, s_2, \dots, s_N)$ on a quantum computer, we employ a basis state encoding:
\begin{equation}
\ket{s} = \bigotimes_{n=1}^N \ket{s_N}, \quad \braket{s_n}{t_n} = \delta_{s_n,t_n}, \quad \braket{\vec{s}}{\vec{t}} = \delta_{\vec{s},\vec{t}}
\end{equation}
on a composite Hilbert space $\mathcal{H} = \bigotimes_{n=1}^N \mathcal{H}_n$, where $\dim(\mathcal{H}_n) = d$ and $\dim(\mathcal{H}) = d^N$.
The loss function $H(\vec{s})$ is then represented by a corresponding diagonal operator, the Hamiltonian $\hat{H}$, which acts on the basis states as follows:
\begin{equation}
\hat{H} \ket{\vec{s}} = H(\vec{s}) \ket{\vec{s}}.
\end{equation}
Apart from the conventional stacking sequence retrieval methods in the comparative study, the algorithms we demonstrate are variational algorithms designed to minimize the expectation value of the loss function:
\begin{equation} 
\ket{\vec{s}_{\text{optimal}}} = \argmin_{\ket{\psi}} \ev{\hat{H}}{\psi}. \end{equation}
The manufacturing constraints are enforced by adding penalty terms to the loss function for each constraint:
\begin{equation} H_{\text{total}}(\vec{s}) = H(\vec{s}) + \sum_{c} H_c(\vec{s}), \end{equation}
where the sum runs over all constraints $c$. Furthermore, each penalty term $H_c$ includes a tunable scaling factor $\gamma_c$.

For a detailed discussion of the basis-state encoding, the loss function definition and the penalty terms for the manufacturing constraints, we refer to our publication \cite{Wulff2024}.

\subsubsection{Algorithms} \label{sec:algos}

In our previous work \cite{Wulff2024}, we demonstrated our approach using two variational quantum algorithms -- QAOA \cite{Farhi2014,Farhi2016} and a hardware-efficient (HWE) approach \cite{Cerezo2021,Endo2021,Bharti2022}  -- as well as DMRG, a classical tensor network-based ground state solver \cite{White1993,Nishino1996,Verstraete2004,Schollwock2011}.
The results were promising, and with this report we aim to extend our demonstrations and address several open questions from the previous study.

\paragraph{DMRG with all constraints:} In our earlier work, while we showed how the listed manufacturing constraints could be implemented as penalties -- in general but also specifically as matrix product operators for DMRG -- the demonstrations only included the disorientation constraint. In this work, building on our previous publication, we have incorporated all of the listed manufacturing constraints in Section~\ref{subsec:constraints} into the matrix product operator that represents the objective function.  While for the constraint case we focus on conventional ply angles in $45^\circ$ intervals, we also performed DMRG without constraints for ply angles with $15^\circ$ separation. More details of the implementation can be found in \cref{SuppClassical}. In particular, the nearest-neighbor coupling is implemented in DMRG to control the clustering and dispersion of plies with the same angle. Details can be found in the supplementary materials \cref{SuppCluster}.

\paragraph{F-VQE:} We observed in our earlier work that, during the HWE optimization, the qubit states tended to collapse into local basis states after just one sweep, effectively eliminating any superposition. Despite this, we found that increasing the number of repetitions in the HWE quantum circuit raised the probability of retrieving the optimal state. This suggests that with more circuit depth and thus increasing initial entanglement, a greater information exchange between the qubits improves the overall optimization. 

To address the issue of immediately collapsing the qubit states, we are now using a more sophisticated quantum algorithm, the Filtering Variational Quantum Eigensolver (F-VQE) \cite{Amaro2022}. Unlike traditional VQE, which typically optimizes the energy expectation value, F-VQE shapes the probability distribution of the variational state according to a filter function. This function increases the probabilities of lower-energy states and suppresses higher-energy ones, effectively filtering out the higher energies. By targeting this modified probability distribution, the algorithm avoids the complete collapse of the quantum state.

We implemented the F-VQE algorithm in Python, following the original publication introducing this approach. We then performed state-vector simulations of the algorithm for laminates with $N=8$ and $N=10$ plies, corresponding to 16 and 20 qubits, using the same hardware-efficient circuit as in \cite{Amaro2022,Wulff2024}. Additionally, we conducted buckling factor maximization based on the biaxial loading buckling factor defined in \cref{eq:lambdaB} in the supplementary material \cref{SuppLaminate}. For this permutation-based search, we employed a parameterized quantum circuit with excitation-preserving gates. Further technical details on the experiments can be found in the supplementary material \cref{SuppTechInfoFvqe}.

\paragraph{Classical algorithms for comparison and scalability analysis:} While we compared the mentioned quantum methods in our previous work, we did not include comparisons with classical approaches currently used for stacking sequence retrieval, particularly when dealt with problems of increasing scales (e.g., number of plies). To fill this gap, which was also mentioned in the feedback sessions, we perform a comparative study with two state-of-the-art libraries: \textit{LAYLA} \cite{layla} (based on Beam-Search algorithm) and \textit{Opti-BLESS} \cite{optibless} (based on Genetic Algorithm). Using DMRG enables us to benchmark our approach with a high ply count -- a scale inaccessible to full state-vector simulations -- and compare it to these established methods. Despite DMRG's strict limitations on entanglement, it offers a conservative baseline, which actual quantum algorithms will likely exceed.

To demonstrate scaling, we tested the three algorithms on symmetric laminates with ply counts ranging from $N=15$ to $N=200$ (half of the full laminate) and conventional ply-angles, i.e. angles in $45^\circ$ intervals. For each ply number $N$, we ran all three algorithms using the same set of target lamination parameters to ensure consistency in the comparison. For DMRG, we evaluated a range of bond dimensions in the matrix product state to assess the tradeoff between accuracy and runtime. In DMRG, constraints were implemented as penalty terms within the objective function, whereas \textit{LAYLA} partially enforces constraints during search tree construction and further applies a repair algorithm to the resulting stacking sequences \cite{Fedon2021,Fedon2021relay}. Although \textit{Opti-BLESS} also offers options for constraint enforcement, we excluded them due to the significantly increased computation time required.

For each tested ply number $N$, we generated a diverse set of target lamination parameters corresponding to valid stacking sequences. This approach ensures that we can directly evaluate the quality of the results based on the value of the loss function, as the optimal solution will always yield a value of 0.

Additionally, we performed DMRG without constraints on a set of allowed ply angles with $15^\circ$ separation, demonstrating that our approach can handle a broader range of ply-angle configurations. For additional technical details and algorithm settings, please refer to the supplementary materials \cref{SuppClassical}.

\subsection{Results}

\paragraph{F-VQE:} With $N=8$ and $N=10$ plies, F-VQE consistently identified the optimal state across all trials. Figures \ref{fig:fvqeres8} and \ref{fig:fvqeres10} in \cref{SuppResults} display, as a function of optimizer iterations, the expectation values of the objective function, the probability of measuring the optimal solution, and the ratio of constraint-compliant states within the superposition -- both with and without an added penalty for enforcing constraints. Individual traces for different target lamination parameters are shown in blue, with their average in black.

Results from the buckling factor maximization are presented in Figures \ref{fig:buckling8} and \ref{fig:buckling10}, where we show histograms of buckling factors obtained from the superposition of states at various optimization stages across trials with different ply-angle counts. In the top half, the histograms show results without constraint enforcement, while the bottom half includes enforced constraints, with constraint-violating states in the superposition highlighted in red. Since the range of achievable buckling factor depends on ply-angle counts, the necessary penalty magnitude varies across trials. Thus, we initially conducted the experiments using a uniformly low penalty, then incrementally increased it for trials that ended in constraint-violating states until optimization produced a valid result, as indicated in the plot.

\paragraph{Comparative study:} \Cref{fig:dmrgconv} displays the results from our comparative study with conventional ply angles and a full set of constraints, showing the accuracy and computational time of DMRG, \textit{LAYLA}, and \textit{Opti-BLESS} across all sampled configurations. DMRG and \textit{LAYLA} only produced valid stacks, with a few rare exceptions for DMRG at simulataneously low ply counts and bond dimensions. \Cref{fig:dmrg15deg} shows results for DMRG with ply angles with $15^\circ$ intervals, without manufacturing constraints. 

\paragraph{Clustering and dispersion:} Using our approach to control clustering and dispersion (\cref{SuppCluster}), \Cref{fig:progressionbias} illustrates how a specific stacking sequence evolves across a range of applied biases: from strong clustering into a few thick plies under negative biases to rapid alternation of ply angles under positive biases.

\subsection{Discussion}

\begin{figure}[h!]
    \centering
    \includegraphics[scale=0.9]{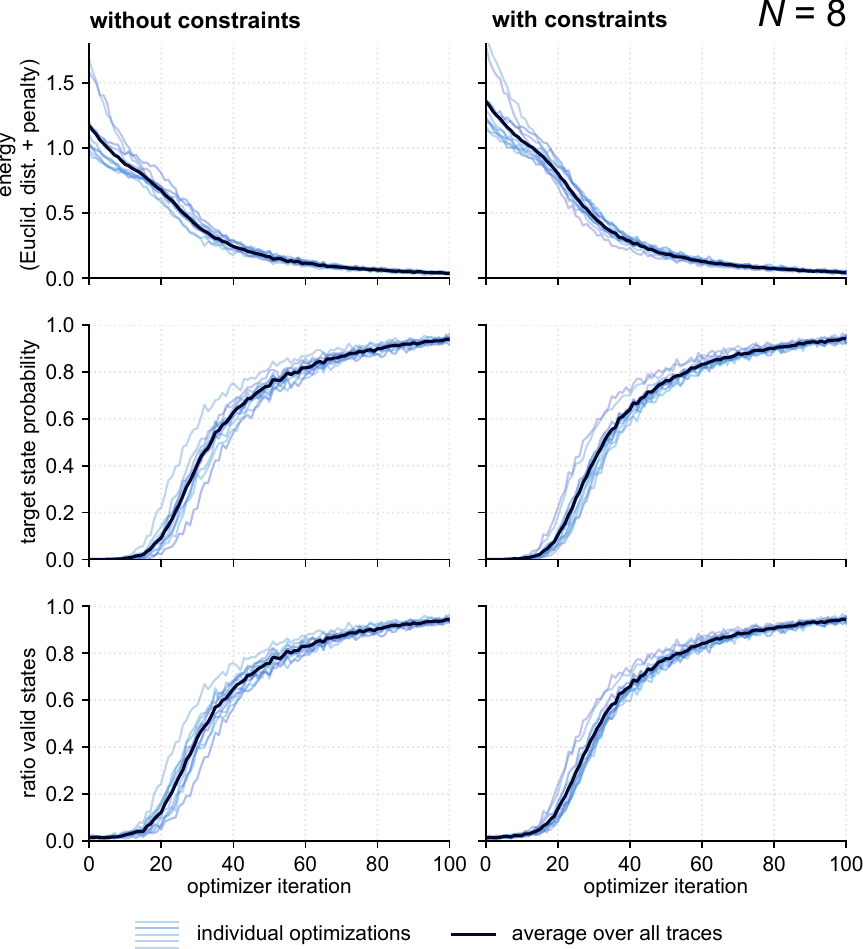}
    \caption{Results of lamination parameter search with F-VQE for $N=8$ plies, with and without constraints. Each blue line represents a single run of the algorithm with varying target lamination parameters. The average is shown in black.}
    \label{fig:fvqeres8}
\end{figure}

\begin{figure}
    \centering
    \includegraphics[scale=0.9]{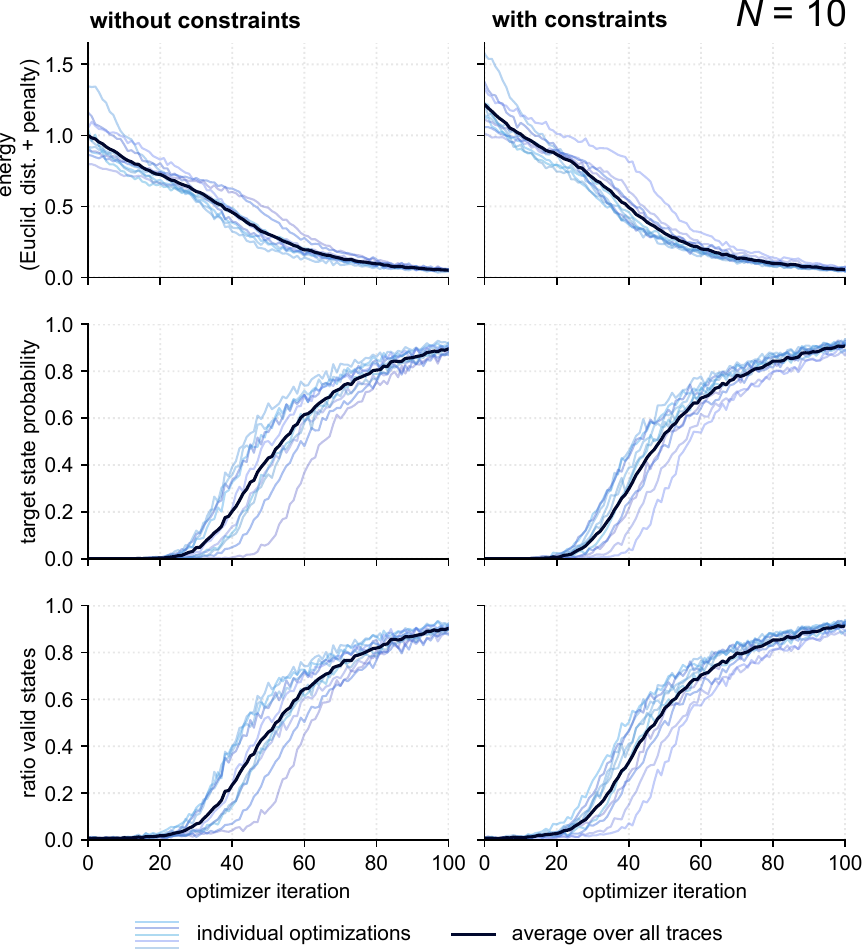}
    \caption{Results of lamination parameter search with F-VQE for $N=10$ plies, with and without constraints. Each blue line represents a single run of the algorithm with varying target lamination parameters. The average is shown in black.}
    \label{fig:fvqeres10}
\end{figure}

\paragraph{F-VQE:} As shown in Figures \ref{fig:fvqeres8} and \ref{fig:fvqeres10}, the search for stacking sequences that match the target lamination parameters found the optimal state reliably for all tested parameters with both $N=8$ and $N=10$ plies. However, caution is advised when generalizing these results to larger ply numbers, as the high number of shots taken during optimization nearly guarantees that the optimal state will appear at some point, which is highly unlikely with significantly more plies. Nevertheless, the results show that F-VQE effectively addresses issues in our previous work noted in \cref{sec:algos}: whereas our prior experiments saw each qubit collapse immediately to a basis state during optimization, limiting paths to the optimal solution, F-VQE maintains a superposition of states throughout, as seen in the ratios for the target state and valid states, which remain between 0 and 1 during the optimization rather than converging fully to either extreme.

Further, F-VQE capitalizes on good states by amplifying their amplitudes, in contrast to the HWE method in our previous work, which only accessed the expectation value of the Hamiltonian and provided limited feedback on the actual state distribution. In HWE, even if a good state appears, it may only slightly improve the expectation value and might not be measured again, whereas F-VQE increases the likelihood of re-measuring these good states by explicitly amplifying them in the distribution. This indicates that F-VQE effectively resolves the concerns raised in \cref{sec:algos} regarding our prior HWE approach.

In terms of constraint enforcement, this experiment offers limited insights since the optimal solution already satisfies the constraints, dominating the trend for valid states overall. However, the figure reveals that adding a penalty does shift the ratio of valid states toward a higher proportion. Constraint effectiveness will be evaluated more conclusively in our buckling maximization experiments, where the optimal state depends significantly on constraint enforcement.

\begin{figure}
    \centering
    \centering
    \begin{adjustbox}{center, max width=\paperwidth}\includegraphics[scale=0.9]{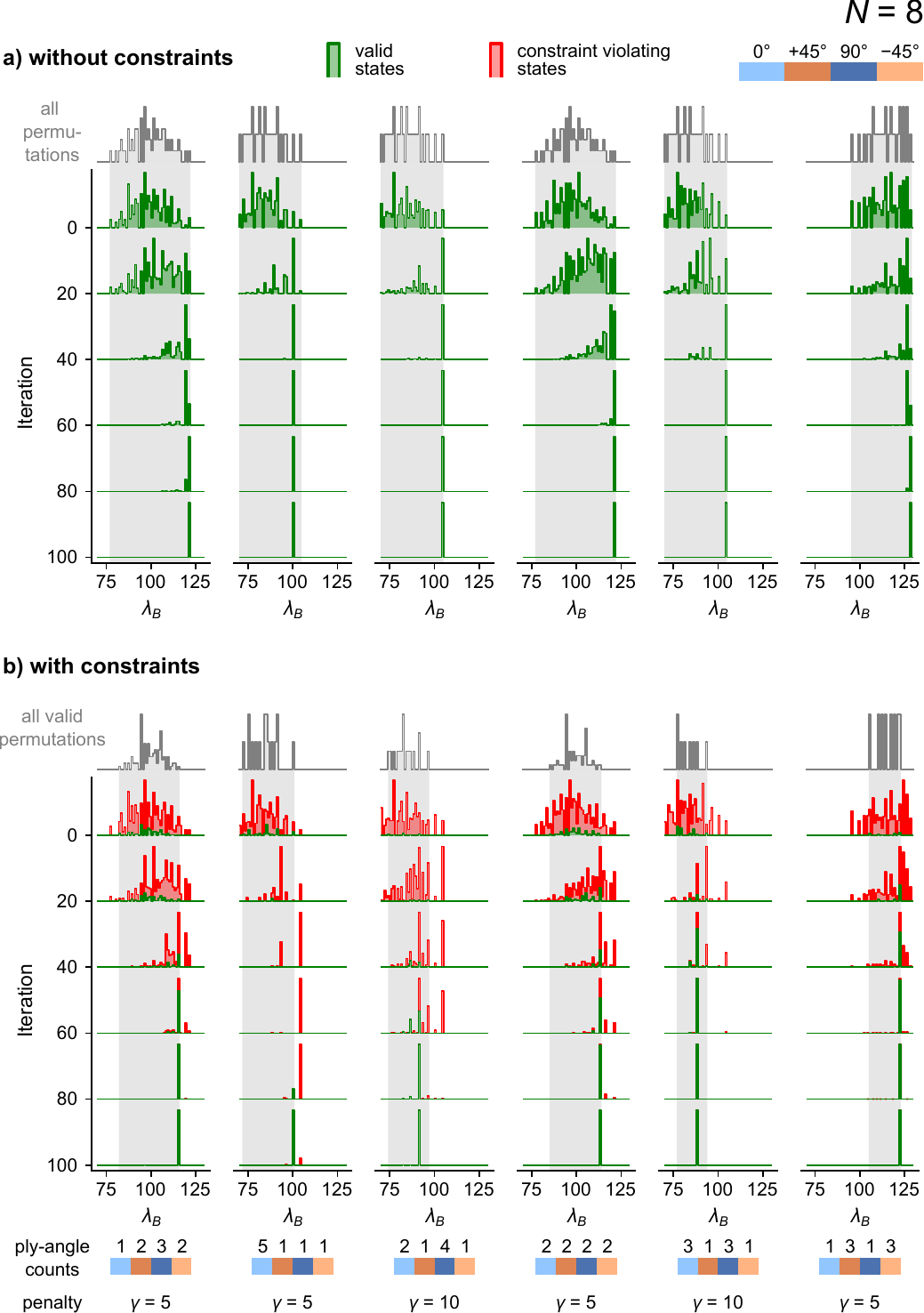}\end{adjustbox}
    \caption{Results of buckling factor maximization with F-VQE for $N=8$ without (a) and with (b) constraints. The top row shows a histogram of all possible permutations, with its range extended to the other rows. In the constraint case, invalid states are shown in red.}
    \label{fig:buckling8}
\end{figure}

\begin{figure}
    \centering
    \begin{adjustbox}{center, max width=\paperwidth}\includegraphics[scale=0.9]{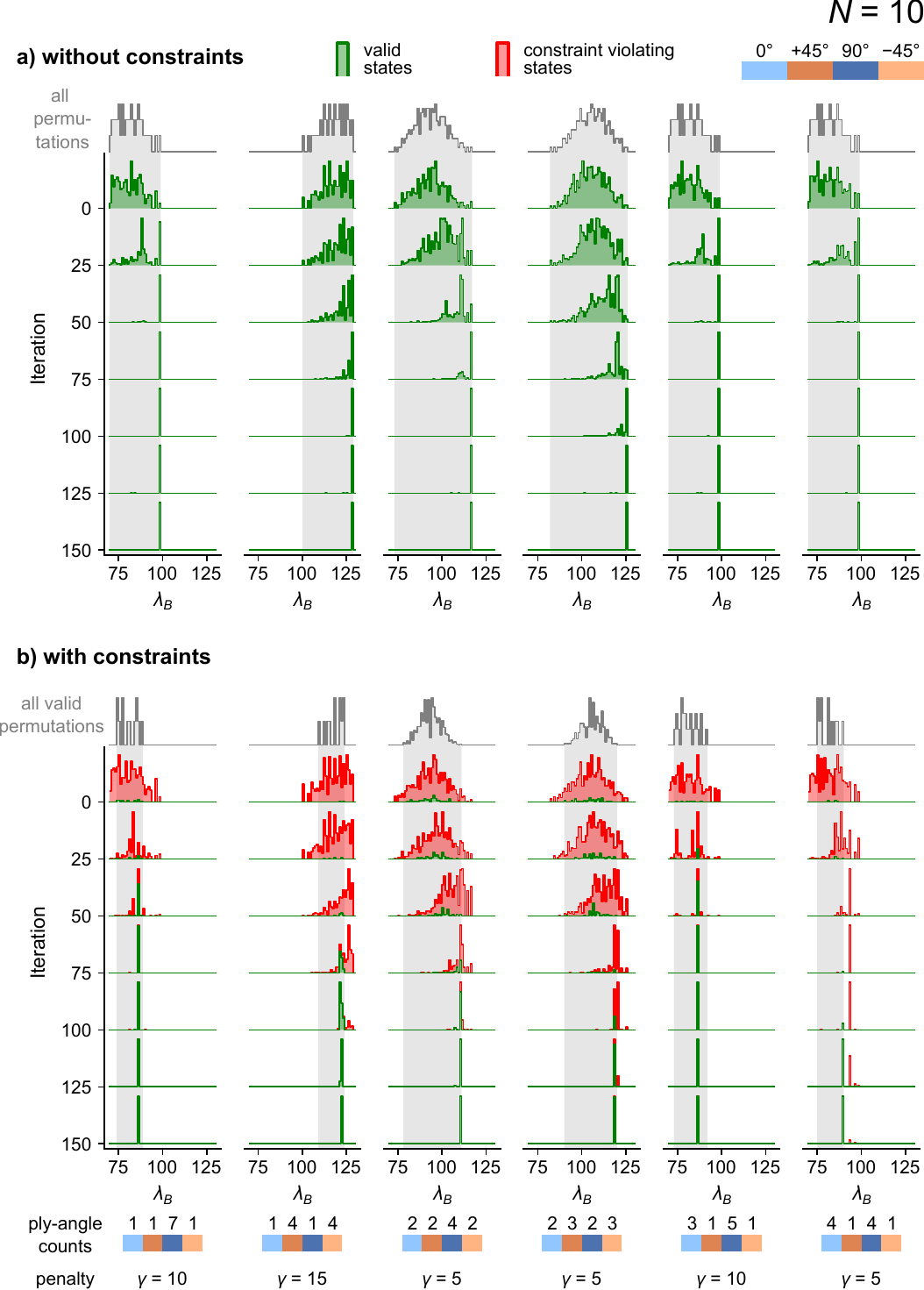}\end{adjustbox}
    \caption{Results of buckling factor maximization with F-VQE for $N=10$ without (a) and with (b) constraints. The top row shows a histogram of all possible permutations, with its range extended to the other rows. In the constraint case, invalid states are shown in red.}
    \label{fig:buckling10}
\end{figure}

For the buckling factor maximization experiment, we can compare histograms of buckling factors (Figures \ref{fig:buckling8} and \ref{fig:buckling10}, green and red) to the theoretical distribution of all possible states (grey, top row) to observe that the algorithm generally identifies a near-optimal solution, albeit with a few slightly suboptimal exceptions. The penalty terms prove effective in enforcing constraints, though it appears to make finding a near-optimal solution more challenging for the algorithm. Notably, the buckling factor objective can be replaced with any other, such as fundamental frequency optimization (e.g., \cref{eq:fundafreq} in \cref{SuppLaminate}) or a combination of multiple objectives, highlighting the flexibility of this approach.

An important consideration regarding buckling factor maximization is that while our simulations use state vectors that precisely conserve ply-angle counts, real quantum devices will introduce errors, potentially resulting in deviations from the required ply-angle counts. Techniques from excitation-conserving variational quantum algorithms in quantum chemistry, such as error mitigation and symmetry verification \cite{Endo2021}, could address these issues, though further investigation is needed to determine their effectiveness in achieving the required accuracy.

Scaling up F-VQE for both lamination parameters search and buckling factor maximization is theoretically straightforward, as adding more plies and qubits presents no immediate technical barriers. However, practical challenges arise with increasing ply numbers, especially in the lamination parameter search, where the high number of shots taken during optimization nearly guarantees that the optimal state will appear at some point in smaller cases like $N=8$ and $N=10$ plies. With significantly more plies, however, this likelihood diminishes. In the buckling maximization case, it will also be valuable to assess whether the partial swaps on qubit pairs (requiring $N(N-1)$ two-qubit gates) are sufficient to maintain high-quality results with significantly more plies. To address these scaling challenges, further investigation is needed into circuit structure, repetition number, and optimization schemes -- such as optimizing parameters in subsets rather than all at once -- to ensure effectiveness at larger scales.

\begin{figure}
    \centering
    \begin{adjustbox}{center, max width=\paperwidth}\includegraphics[scale=1.0]{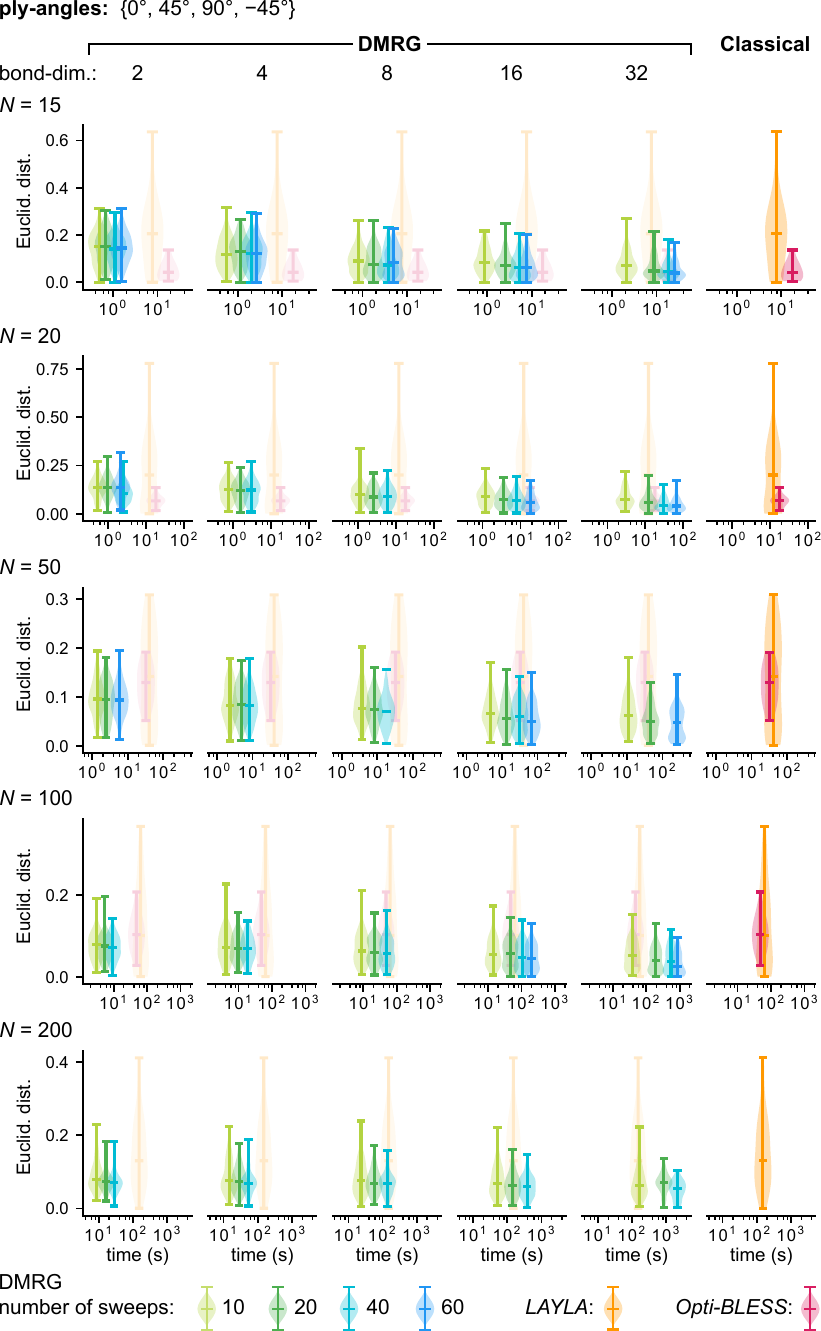}\end{adjustbox}
    \caption{Results for DMRG with conventional ply-angles and manufacturing constraints, in comparison to \textit{LAYLA} and \textit{Opti-BLESS}. Shown are the distribution of the resulting distances to the target parameters for each number of performed sweeps. The middle horizontal line displaying the median of the distribution, while the vertical line is positioned at the mean runtime. The results from the classical algorithms are included in the background of the plots for easier comparison.}
    \label{fig:dmrgconv}
\end{figure}

\paragraph{DMRG in comparison to conventional methods:} In the case of conventional ply angles (\cref{fig:dmrgconv}), DMRG generally outperformed \textit{LAYLA} in accuracy and required less computation time across nearly all configurations tested. Constraint violations with DMRG were minimal, occurring only in rare cases with low ply counts and bond dimensions, indicating that the penalties were well-tuned. \textit{LAYLA} similarly showed no constraint violations, owing to the repair algorithm it employs \cite{Fedon2021relay}. Notably, \textit{LAYLA} includes a large number of hyperparameters, which may improve performance with expert tuning. By contrast, DMRG is relatively straightforward to tune, relying primarily on bond dimension, the number of sweeps, and penalty strength as its main hyperparameters. The bond dimension and sweep count were analyzed in detail (see \cref{fig:dmrgconv}), and setting the penalties inversely proportional to the ply count $N$ proved to be effective (see \cref{tab:dmrg_configurations}).

\textit{Opti-BLESS}, on the other hand, performed comparably to DMRG in terms of accuracy, with a decrease in performance relative to DMRG as the ply count $N$ increased. In terms of runtime, \textit{Opti-BLESS} was generally similar to \textit{LAYLA} but showed longer durations for lower ply counts. It should be noted that due to time limitations, \textit{Opti-BLESS} was tested without constraint enforcement, which limits the comparability of these results. Adding constraints will increase runtime significantly and may negatively impact accuracy. 

\begin{figure}
    \centering
    \begin{adjustbox}{center, max width=\paperwidth}\includegraphics[scale=1.0]{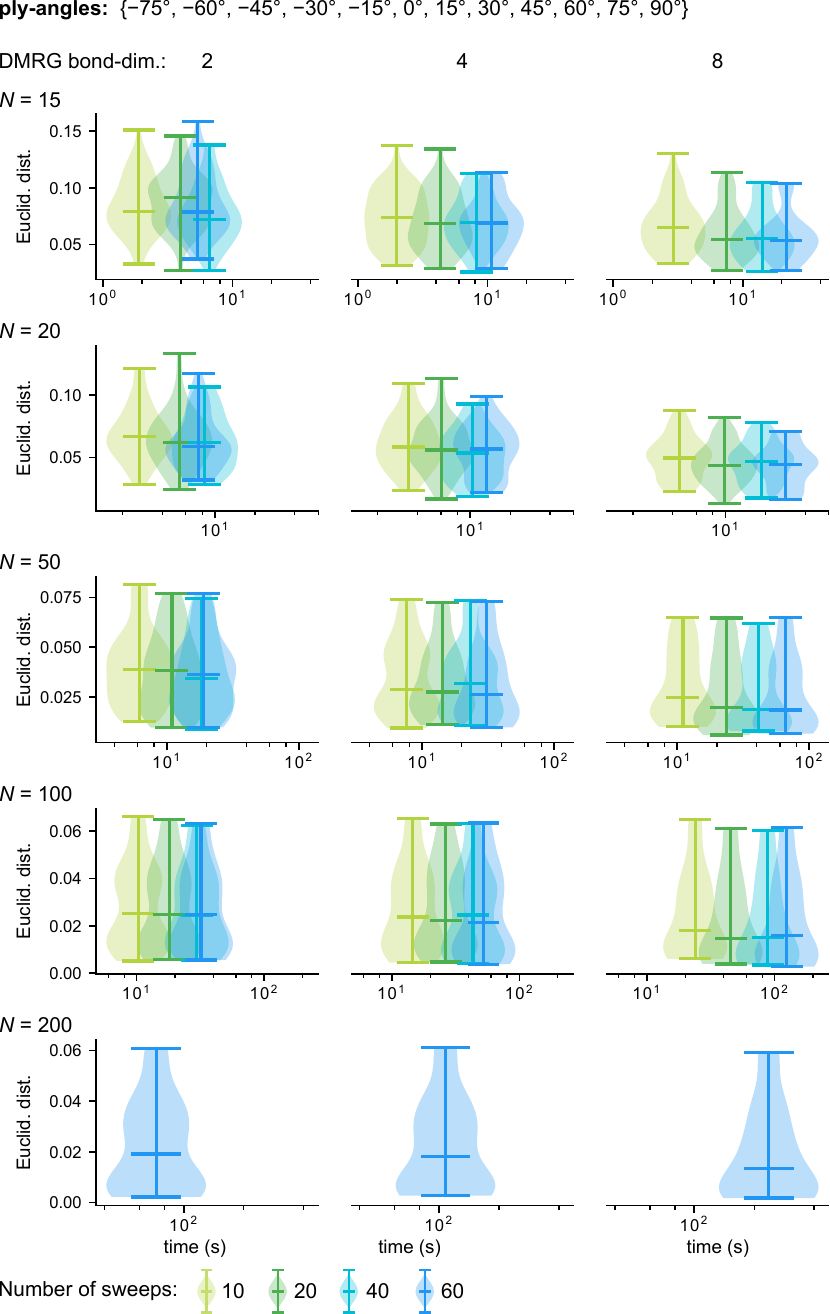}\end{adjustbox}
    \caption{Results of DMRG with 15 degree separation in varying configurations. Shown are the distribution of the resulting distances to the target parameters for each number of performed sweeps. The middle horizontal line displaying the median of the distribution, while the vertical line is positioned at the mean runtime.}
    \label{fig:dmrg15deg}
\end{figure}

In terms of scalability, the DMRG results in \cref{fig:dmrgconv} show that the error remains stable around 0.2 or less as the number of plies increase from 15 to 200. The computational time for the same number of sweeps stays mostly stable with respect to the number of plies, except for large number of plies (100 to 200) where a ten-fold increase in time can be observed for 40 and 60 sweeps with bond dimensions of 16 and 32. In addition, \cref{fig:dmrg15deg} shows that DMRG can effectively handle ply angles at $15^\circ$ separation, which is a huge increase in ply-angle choices over the conventional $45^\circ$ separation. Due to time limitation, these trials were run without the constraints. Nevertheless, the data serves to show the scalability of the method with unconventional ply angles. Both the accuracy and computational time stay stable as the number of plies increase from 15 to 200. These results highlight that our approach scales well both in the number of plies and in the choices of ply angles, broadening its applicability in diverse stacking sequence retrieval tasks.

In summary, these results indicate that DMRG, as a quantum-inspired approach to stacking sequence retrieval, demonstrates strong performance, even without the full entanglement capabilities that a quantum computer could provide.

\begin{figure}
    \centering
    \includegraphics[scale=0.9]{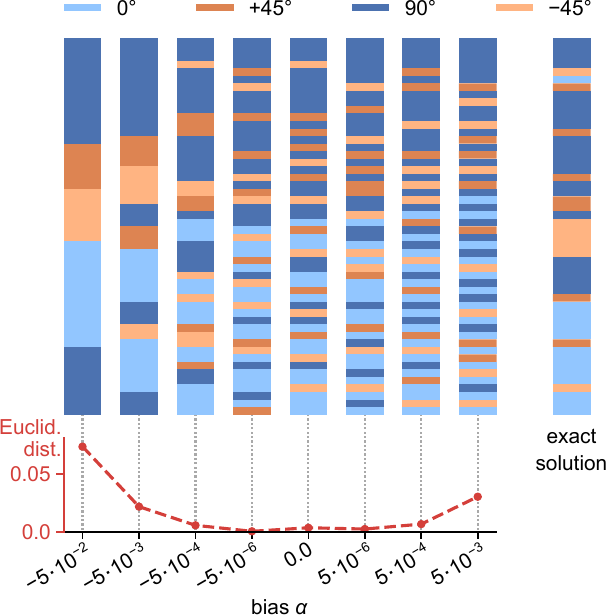}
    \caption{Effect of an added bias $\alpha$ on nearest neighbor interactions, on a laminate of $N=50$ plies. For each stack, we chose the best of 10 independent trials of DMRG at bond-dimension 32.}
    \label{fig:progressionbias}
    \vspace{7cm}
\end{figure}

\paragraph{Clustering and dispersion experiment:} To control the clustering and dispersion of same-angle plies, \Cref{fig:progressionbias} demonstrates the effectiveness of adding a bias $\alpha$ as a nearest-neighbor interaction term. For large negative values of $\alpha$, we observe the formation of a few thick plies, while large positive values lead to almost completely alternating ply angles. In the middle range, where the bias is noticeable enough to impact the distribution, it has only a minor effect on the accuracy, comparable to the natural variation in the Euclidean distance $\mathrm{dist}_{\vec{\xi}}$ observed without any bias. For high values of $\alpha$, however, the distance increases, as expected, since a fully clustered or fully dispersed stacking sequence is less likely to match the target lamination parameters closely. While \Cref{fig:progressionbias} shows one example, this behavior was consistent across our tests. Notably, for $\alpha > 0$, we frequently observed larger blocks of same-angle plies forming toward the outer regions of the stack, likely because the outer plies contribute more significantly to the lamination parameters than the inner ones. To mitigate this effect, a possible approach is to make the bias dependent on the ply index $n = 1,2,\dots,N$, applying a higher bias at the outer plies to ensure a uniform effect across the sequence. Overall, this demonstrates that cost-reduction measures in terms of encouraging thicker plies, can be seamlessly integrated into our approach.

\clearpage
\section{Conclusion and Outlook}

In this work, we have proposed the stacking sequence retrieval problem, the second step in the bi-level optimization of composite laminates, as the ``Golden Application'' of quantum computing for the quantum mobility quest. We opened with an overview over the bi-level optimization approach and our motivation to this challenge. We then presented several novel extensions to our submission in Phase 1 of the challenge, where we have included all the manufacturing constraints in the DMRG algorithm, introduced the clustering-dispersion constraint for variable-thickness ply-blocks and a novel objective function on buckling factor, and implemented F-VQE as a more powerful quantum solver than the original VQA. In addition, we have compared our DMRG algorithm with two widely-used classical algorithms on a range of different test cases at different thickness scales.

Our results strongly support the validity of our approach. First, we demonstrated how sophisticated algorithms like F-VQE effectively address limitations we observed with standard VQA in our previous work \cite{Wulff2024}. Second, the advantage of our method was evident in large-scale DMRG tests. With the full set of constraints, our results were competitive against the state-of-the-art conventional stacking sequence retrieval methods, outperforming them on most cases in both accuracy and efficiency. Our quantum-inspired DMRG algorithm showed excellent scalability with both increasing number of plies and increasing ply-angle choices. It should be noted, however, that DMRG is significantly restricted in its capacity to capture the level of entanglement achievable by true quantum algorithms. Given F-VQE’s improved performance and ongoing advancements in quantum algorithms, we are optimistic that leveraging an actual quantum device, instead of simulated entanglement, could ultimately outperform the classical and our current DMRG methods.

Further, we demonstrated the flexibility of our approach which not only allows for lamination parameter optimization, but also extends to alternative objectives such as buckling factor maximization, and even cost-reduction strategies through the control of clustering and dispersion of same-angle plies to have variable-thickness ply-blocks. Together with the high number of plies that we can handle in our algorithm, this flexibility allows designers a fine control over the stacking sequence design space to achieve the desired balance between different objectives. 

The innovative way of interpreting an old problem with quantum computing narrative and approach would be only the beginning of tapping into the great potential held by the future. Many materials design challenges, such as functional graded multi-material system design, could benefit from the approach presented here. For example, our algorithm can be readily employed in the hybrid-material optimization framework at the German Aerospace Center (DLR) \cite{Silva2020}. The retrieved stacking sequence from our algorithm can be directly fed to a suitable nonlinear finite element software to build high-fidelity layer-wise models of the laminate to predict accurately its damage evolution and strength \cite{chen2014floating,chen2016modelling,chen2017modelling}. Fast connection between the high-fidelity nonlinear finite element model and the bi-level optimization framework can be achieved through machine-learning surrogate of the finite element model, thereby providing in-situ feedback to the designers in this optimization loop. In the future, we will further extend our algorithm for the optimization across multiple panels to retrieve the stacking sequence that obeys the blending constraints \cite{Bordogna2020}. Our algorithm is not limited to designing composites in vehicle structures -- many components in the renewable energy sector, such as wind turbines, batteries, and solar panels, exhibit a layer-wise structure for which a ``stacking sequence'' of design choices would need to be made. We therefore envision the generalization of our algorithm for these optimization problems. Moreover, as we explore more types of penalty functions, it could even lead us into designing materials on the atomic level, for example, MAX phase with varying constituent atoms across layers, high entropy ultra-high temperature ceramics or metals alike.

\clearpage
\section{Bibliography}
\printbibliography[heading=none]

\clearpage
\section{Supplementary information}
\label{Supp}

\subsection{Additional information for the bi-level laminate optimization} \label{SuppLaminate}

In this section, we provide additional details on the bi-level optimization and stacking sequence retrieval. For a more comprehensive explanation, we refer to our previous publication \cite{Wulff2024}.

In laminated composites, the orientation of the individual plies dictates the mechanical behavior of the material. To a linear approximation, this behavior is commonly characterized by the stiffness or $\mathbf{ABD}$ matrix, which relates the stress and moment resultants $\vec{N}$ and $\vec{M}$ to the in-plane and out-of-plane deformations $\vec{\varepsilon}^{: 0}$ and $\vec{\kappa}$ \cite{Reissner1961,Dong1962,Tsai1980}:
\begin{equation} \label{eq:hook_suppl}
    \begin{bmatrix}
    \vec{N} \\ \vec{M}
\end{bmatrix} = \begin{bmatrix}
    \mathbf{A} & \mathbf{B} \\ \mathbf{B} & \mathbf{D}
\end{bmatrix} \begin{bmatrix}
    \vec{\varepsilon}^{\: 0} \\ \vec{\kappa}
\end{bmatrix}.
\end{equation}
The stiffness matrix is determined by the laminate thickness $h$, the intrinsic material properties of a single ply, and the ply angle sequence $\vec{\phi} = (\phi_1, \phi_2, \dots, \phi_N)$. Using the lamination parameters \cite{Tsai1968,Miki1991,IJsselmuiden2011,Albazzan2019}:
\begin{align} \label{eq:lp}
    (v_1^A,v_2^A,v_3^A,v_4^A) &= \frac{1}{N} \sum_{n=1}^N \qty(z_n - z_{n-1}) \qty(\cos(2\phi_n),\sin(2\phi_n),\cos(4\phi_n),\sin(4\phi_n)),\\
     (v_1^B,v_2^B,v_3^B,v_4^B) &= \frac{2}{N^2} \sum_{n=1}^N \qty(z_n^2 - z_{n-1}^2) \qty(\cos(2\phi_n),\sin(2\phi_n),\cos(4\phi_n),\sin(4\phi_n)), \notag \\
      (v_1^D,v_2^D,v_3^D,v_4^D) &= \frac{4}{N^3} \sum_{n=1}^N \qty(z_n^3 - z_{n-1}^3) \qty(\cos(2\phi_n),\sin(2\phi_n),\cos(4\phi_n),\sin(4\phi_n)), \notag
\end{align}
where the distance of ply $n$ from the midplane captured in \(z_n = -N/2 + n\), the stiffness matrix can be written in the following compact form:
\begin{align} \label{eq:abd_from_lp}
\mathbf{A} &= h \qty(\bm{\Gamma}_0 + \bm{\Gamma}_1 v_1^A + \bm{\Gamma}_2 v_2^A + \bm{\Gamma}_3 v_3^A + \bm{\Gamma}_4 v_4^A),\\
\mathbf{B} &= \frac{h^2}{4} \qty(\bm{\Gamma}_0 + \bm{\Gamma}_1 v_1^B + \bm{\Gamma}_2 v_2^B + \bm{\Gamma}_3 v_3^B + \bm{\Gamma}_4 v_4^B), \notag\\
\mathbf{D} &= \frac{h^3}{12} \qty(\bm{\Gamma}_0 + \bm{\Gamma}_1 v_1^D + \bm{\Gamma}_2 v_2^D + \bm{\Gamma}_3 v_3^D + \bm{\Gamma}_4 v_4^D). \notag
\end{align}
Here the intrinsic material properties are expressed via the matrices $\bm{\Gamma}_l$, which contain the Tsai-Pagano material invariants $U_1, \dots, U_5 \in \R$ \cite{Tsai1968,IJsselmuiden2011,Bordogna2020}:
\begin{equation} \label{eq:gammas}
    \bm{\Gamma}_0 = \begin{bmatrix}
    U_1 & U_4 & 0  \\
    U_4 & U_1 & 0 \\
    0 & 0 & U_5
\end{bmatrix}, \quad \bm{\Gamma}_1 = \begin{bmatrix}
    U_2 & 0 & 0  \\
    0 & -U_2 & 0 \\
    0 & 0 & 0
\end{bmatrix}, \quad \bm{\Gamma}_2 = \begin{bmatrix}
    0 & 0 & U_2/2 \\
    0 & 0 & U_2/2 \\
    U_2/2 & U_2/2 & 0 
\end{bmatrix}
\end{equation}
\begin{equation*}
    \bm{\Gamma}_3 = \begin{bmatrix}
    U_3 & -U_3 & 0  \\
    -U_3 & U_3 & 0 \\
    0 & 0 & -U_3
\end{bmatrix}, \quad \bm{\Gamma}_4 = \begin{bmatrix}
     0 & 0 & U_3 \\
     0 & 0 & -U_3 \\
    U_3 & -U_3 & 0
\end{bmatrix}.
\end{equation*}

In this work, we focus on ply-angle sequences that are symmetric around the midplane of the material. In such cases, the $B$-lamination parameters and the $\mathbf{B}$-matrix vanish. By indexing only one half of the laminate from the midplane outward and set $N$ to half the ply number, the lamination parameters take the form:
\begin{align} \label{eq:lpsym}
    (v^A_1,v^A_2,v^A_3,v^A_4) &= \frac{1}{N} \sum_{n=1}^N (z_n - z_{n-1}), \qty(\cos(2\theta_n),\sin(2\theta_n),\cos(4\theta_n),\sin(4\theta_n)), \\
    (v^D_1,v^D_2,v^D_3,v^D_4) &= \frac{1}{N^3}\sum_{k=1}^N (z_n^3 - z_{n-1}^3) \qty(\cos(2\theta_n),\sin(2\theta_n),\cos(4\theta_n),\sin(4\theta_n)), \notag\\
    z_n &= n \notag.
\end{align}

While the target lamination parameters for conventional stacking sequence retrieval depend solely on the ply angles and ply number, the buckling factor is determined by the elements of the stiffness matrix. In this work, we use the closed-form solution for bi-axial loading of a simply supported plate (see, e.g., \cite{Bordogna2020}):
\begin{equation} \label{eq:lambdaB}
\lambda_B = \pi^2 \frac{D_{11} (m/a)^4 + 2(D_{12} + 2 D_{33}) (m/a)^2 (n/b)^2 + D_{22}(n/b)^4}{(m/a)^2 N_X + (n/b)^2 N_Y},
\end{equation}
where buckling occurs for a plate with dimensions $a$ and $b$ when $0 < \lambda_B < 1$, under the stresses $N_X$ and $N_Y$ in the in-plane directions, with corresponding half-wave numbers $n$ and $m$.
According to equations (\ref{eq:abd_from_lp}) and (\ref{eq:gammas}), the elements of the $D$ matrix are derived from the Tsai-Pagano invariants $U_1, U_2, U_3, U_4, U_5$, which are calculated via the $\mathbf{Q}$ matrix as follows \cite{Tsai1980,IJsselmuiden2011}:
\begin{align} 
U_1 &= \frac{1}{8}\qty(3 Q_{11} + 3 Q_{22} + 2 Q_{12} + 4 Q_{66}), \\
U_2 &= \frac{1}{2}\qty(Q_{11} - Q_{22}), \\
U_3 &= \frac{1}{8}\qty(Q_{11} + Q_{22} - 2 Q_{12} - 4 Q_{66}), \\
U_4 &= \frac{1}{8}\qty(Q_{11} + Q_{22} + 6 Q_{12} - 4 Q_{66}), \\
U_5 &= \frac{1}{8}\qty(Q_{11} + Q_{22} - 2 Q_{12} + Q_{66}).
\end{align}
These invariants, in turn, depend on the material properties: the longitudinal and transverse Young’s moduli $E_1$ and $E_2$, the shear modulus $G_{12}$, and Poisson’s ratio $\nu_{12}$:
\begin{align}
Q_{11} &= \frac{E_1}{1 - \nu_{12} \nu_{21}}, \\
Q_{12} &= \frac{\nu_{12} E_1}{1 - \nu_{12} \nu_{21}}, \\
Q_{22} &= \frac{E_2}{1 - \nu_{12} \nu_{21}}, \\
Q_{66} &= G_{12}, \\
\frac{\nu_{21}}{\nu_{12}} &= \frac{E_1}{E_2}.
\end{align}
For our demonstration, we used material values from \cite{Bordogna2020}:
\begin{equation}
E_1 = 177 \ \text{GPa}, \quad E_2 = 10.8 \ \text{GPa}, \quad G_{12} = 7.6 \ \text{GPa}, \quad \nu_{12} = 0.27.
\end{equation}
We selected a unit plate with dimensions $a = b = 1$ and set the forces such that $N_X = 2$ and $N_Y = 1$, making the force in the $X$ direction twice as large as in the $Y$ direction. For this demonstration, we focused on the case where $m = n = 1$. To avoid overflow in the exponential filter while keeping the penalty terms balanced, we used the laminate thickness $h$ to adjust the magnitude of the buckling factor. After testing, we found that $h = 0.1$ provided a suitable balance.

Our approach is general enough to optimize objectives beyond buckling; any property that relies on the ABD matrix can be used as the optimization target. One example is maximizing the fundamental frequency $\omega_{mn}$ of the laminate, which depends on elements of the stiffness matrix as follows \cite{Ghasemi2013}:
\begin{equation} \label{eq:fundafreq}
\omega_{m n}^{2}=\frac{L_{13} K_{A}+L_{23} K_{B}+L_{33}}{\rho h},
\end{equation}
where
\begin{equation}
K_{A} =\frac{L_{12} L_{23}-L_{13} L_{22}}{L_{11} L_{22}-L_{12}^{2}}, \qquad K_{B}=\frac{L_{12} L_{13}-L_{11} L_{23}}{L_{11} L_{22}-L_{12}^{2}}
\end{equation}
and
\begin{align}
& L_{11}=D_{11}\left(\frac{m \pi}{a}\right)^{2}+D_{66}\left(\frac{n \pi}{b}\right)^{2} \\
& L_{12}=L_{21}=\left(D_{12}+D_{66}\right)\left(\frac{m \pi}{a}\right)\left(\frac{n \pi}{b}\right) \\
& L_{13}=L_{31}=k_{\mathrm{sh}} A_{55}\left(\frac{m \pi}{a}\right) \\
& L_{22}=D_{66}\left(\frac{m \pi}{a}\right)^{2}+D_{22}\left(\frac{n \pi}{b}\right)^{2}+k_{\mathrm{sh}} A_{55} \\
& L_{23}=L_{32}=k_{\mathrm{sh}} A_{44}\left(\frac{n \pi}{b}\right) \\
& L_{33}=k_{\mathrm{sh}} A_{55}\left(\frac{m \pi}{a}\right)^{2}+k_{\mathrm{sh}} A_{44}\left(\frac{n \pi}{b}\right)^{2}.
\end{align}
Note that these equations employ an extended model compared to \cref{eq:hook}, incorporating transverse-shear stiffness terms, $A_{44}$ and $A_{55}$. While the lamination parameters are not directly applicable, the laminate thickness can still be factored out, allowing the stiffness terms to be calculated from the stacking sequence via the respective $Q_{ij}^n$ matrices for the individual plies $n$:
\begin{equation}
    A_{i j} = \frac{h}{N} \sum_{n=1}^N Q_{i j}^{n}.
\end{equation}

\FloatBarrier

\subsection{Additional information on the algorithms}
\label{SuppTechInfo}

\subsubsection{Additional information on F-VQE} \label{SuppTechInfoFvqe}

\begin{table}[ht]
    \centering
    \label{tab:fvqesetting}
    \begin{tabular}{@{}ll@{}}
        \toprule
        \textbf{Parameter}                     & \textbf{Value}                              \\ 
        \midrule
        Shots & 1000 \\
        Target gradient norm $g_c$ & \\
        $\quad$ LP search & 0.25 \\
        $\quad$ Buckling maximization & 0.1\\
        Learning rate $\eta$ & 1. \\
        Maximum filtering parameter $\tau_\mathrm{max}$ & 200 \\
        Increment of $\tau$ & $0.1 \cdot (\text{last $\tau$})$\\ $\quad$ First iteration & $0.1$\\
        \bottomrule
    \end{tabular}
    \caption{Settings for F-VQE}
\end{table}

\begin{figure}
    \centering
    \includegraphics[scale=1.0]{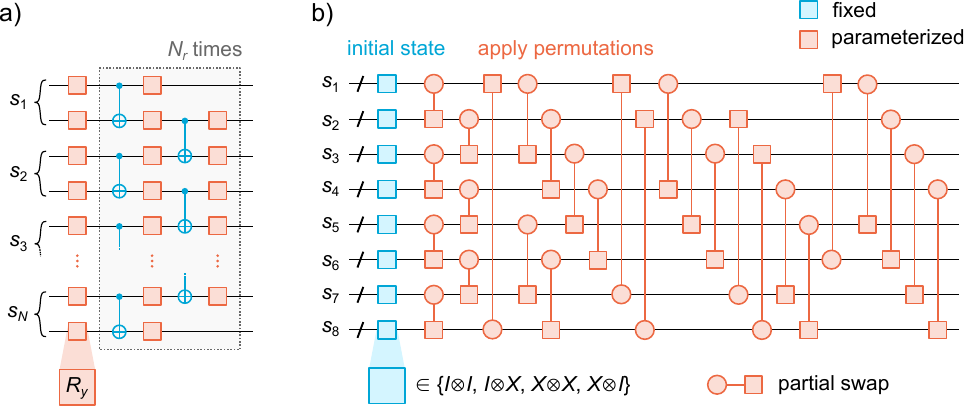}
    \caption{a) The parameterized quantum circuit used lamination parameter search b) The parameterized quantum circuit used for the buckling factor maximization for $N=8$. Fixed gates are shown in blue while parameterized gates are colored orange. As the sign of the partial swap gate (\cref{eq:partialswap}) depends on the order of the ply states, we use a circle to denote and first and a square to denote the second ply to which the gate is applied.}
    \label{fig:vqcs}
\end{figure}

We employ a modified binary encoding to encode the states $\ket{1}$, $\ket{2}$, $\ket{3}$ and $\ket{4}$ corresponding to the conventional ply angles $0^\circ$, $+45^\circ$, $90^\circ$ and $-45^\circ$ \cite{Wulff2024} :
\begin{equation}
    \ket{1} \equiv \ket{00}, \quad, \ket{2} \equiv \ket{01}, \quad, \ket{3} \equiv \ket{11}, \quad, \ket{4} \equiv \ket{10}.
\end{equation}
We used a NumPy-based state-vector simulation of the hardware-efficient quantum circuit, as shown in \Cref{fig:vqcs} a, with $N_\mathrm{rep} = 2$ repetitions. At the start of the optimization, all parameters are set to $0$ except for the last $R_y$ rotation on each qubit, which are set to $\pi/2$ to produce an equal superposition of all basis states. To mimic measurements on real quantum devices, we sampled the probability distribution of the final state-vector 1000 times per evaluation. As described in the F-VQE paper \cite{Amaro2022}, at each iteration, we measured the circuit by shifting each parameter individually. We experimented with different configurations on which parameters are optimized at each iteration, but for $N=8$ and $N=10$ we could not find a significant dependence of the performance on the optimization scheme. We thus only present the results where all parameters are updated at each iteration. The gradient of the filtering operator was then determined in post-processing from the measured counts, ensuring that the gradient remained approximately constant in each iteration, if possible. For lamination parameter search, we employed the inverse filter function to more aggressively amplify low-energy states, with a small constant of $0.001$ added to avoid division by zero. Since, due to the non-polynomial filtering, any operator evaluations were performed in post-processing, we used the Euclidean distance (\cref{eq:euclid}) in the loss function. Constraint in the LP search were enforced with a penalty of $\gamma = 0.05$ for all manufacturing constraint.

For the buckling factor maximization, we selected the negative buckling factor $-\lambda_B$ (see \cref{eq:lambdaB} in \cref{SuppLaminate}) as the objective function to minimize, adding a sufficiently large constant $\lambda_{\mathrm{max}}$ to ensure positivity across all values:
\begin{equation}
    H(\vec{s}) = - \lambda_B(\vec{s}) + \lambda_{\mathrm{max}} + H_{\mathrm{penalties}}
\end{equation}
Furthermore, we provided the algorithm with ply-angle counts that define the allowed permutations and satisfy the balanced condition and 10\% rule. Initial tests enforcing ply-angle counts with penalty terms did not yield the desired results, likely due to the penalty terms overshadowing the buckling factor objective. Therefore, we opted to enforce ply-angle counts directly within the parameterized quantum circuit: we initialized the qubits in a basis state with the correct ply-angle counts and applied quantum gates that perform partial swap operations on the ply states $\ket{st}, \ s,t \in \{1,2,\dots,d\}$:
\begin{equation} \label{eq:partialswap}
    U(\alpha) \ket{st} = \begin{cases}
        \cos(\frac{\alpha}{2}) \ket{st} + \sin(\frac{\alpha}{2}) \ket{ts}, & \text{if $s<t$}\\
        \cos(\frac{\alpha}{2}) \ket{st} - \sin(\frac{\alpha}{2}) \ket{ts}, & \text{if $s>t$}\\
        \ket{st}, & \text{if $s=t$}\\
    \end{cases}
\end{equation}
This gate is inspired by the excitation conserving interactions such as in \cite{Barkoutsos2018} and effectively only create superpositions of permutations of the initial state, keeping the ply-angle counts intact. The complete quantum circuit for $N=8$ is shown in \Cref{fig:vqcs} b. In order to start the optimization in a broad superposition of basis states, we initialize all parameters to $\pi/2$. Unlike in the lamination parameter search, the minimum energy varies between problem instances. To ensure consistent filtering across all trials, we selected the exponential filter as the filter function for this case. In the constrained case, we initially set the penalty to $\gamma = 5$ and for invalid results, we increased the penalty in increments of 5 until a valid stacking sequence was produced.

\subsubsection{Additional information on DMRG and the classical methods} \label{SuppClassical}

\begin{table}[ht]
    \centering
    \label{tab:algorithm_settings_constraints}
    \begin{tabular}{@{}ll@{}}
        \toprule
        \textbf{Constraint}     & \textbf{Penalty}       \\ 
        \midrule
        Disorientation constraint & $1.0/N$                                       \\
        Contiguity constraint & $0.5/N$                                       \\
        10\%-rule  & $0.2/N$                                        \\
        Balanced condition & $0.2/N$                                       \\
        \bottomrule
    \end{tabular}
    \caption{Penalty factors $\gamma$ used in DMRG}
\end{table}
\begin{table}[ht]
    \centering
    \label{tab:dmrg_configurations}
    \begin{tabular}{@{}ll@{}}
        \toprule
        \textbf{Parameter}                        & \textbf{Values}                                                                                           \\ 
        \midrule
        Maximum bond Dimension                 & 2, 4, 8, 16, 32 \\
        Sweeping direction                          & Inward (from $N$ to $1$)                                                                             \\
        Number of sweeps                          & 10, 20, 40, 60                                                                                          \\
        \bottomrule
    \end{tabular}
    \caption{Hyperparameters for our demonstration of DMRG}
\end{table}

We used DMRG with an inward sweeping direction, as our prior work identified this as the most effective among the configurations investigated \cite{Wulff2024}. For DMRG, we employed the square Euclidean distance (\cref{eq:euclid2}) as the loss function due to its efficient matrix product operator representation \cite{Wulff2024}. Each problem instance was tested across multiple maximum bond dimensions, ranging from 2 to 32. Our implementation of DMRG includes a procedure to reduce the bond dimension to towards a basis state, to obtain a single stacking sequence as a result \cite{Wulff2024}. We implemented the tensor networks and performed DMRG using the \textit{ITensors.jl} library in \textit{Julia}. DMRG was applied to configurations with conventional ply angles including manufacturing constraints, as well as to configurations with ply angles in $15^\circ$ intervals without constraints.

For the comparative study, we used \textit{LAYLA} \cite{layla,Fedon2021}, based on a beam search method with a repair strategy \cite{Fedon2021relay}, and \textit{Opti-BLESS} \cite{optibless}, which applies genetic algorithms for stacking sequence retrieval. Each algorithm was run with its standard setup for problem instances, except that \textit{LAYLA}’s tree-search hyperparameters were adjusted to match those in \cite{Fedon2021}. While constraints were included in \textit{LAYLA}, we ran \textit{Opti-BLESS} without constraints, as the resulting runtime increases with constraints made it impractical to obtain sufficient data for a detailed comparison.

To ensure comparability, we used the same set of 40 target lamination parameters for each ply count $N$; however, due to time constraints, some configurations do not include all targets. In particular, the number of trials per configuration for \textit{Opti-BLESS} was limited to 10 different targets.

\subsubsection{Controlling clustering and dispersion of same-angle plies with nearest-neighbor interactions} \label{SuppCluster}

The contiguity constraint limits the number of consecutive plies with the same angle. However, in discussions with Airbus and BMW engineers, we learned that, in some cases, thicker blocks of plies with the same angle may be more desirable. For instance, manufacturing thicker plies with the same orientation can reduce costs compared to producing numerous thin plies with varied orientations.

To explore this, we investigated how ply clustering or dispersion can be controlled by introducing a small bias in a nearest-neighbor coupling term:
\begin{equation}
    H_\alpha (\vec{s}) = \sum_{n=1}^{N-1} \alpha \delta_{s_n,s_{n+1}},
\end{equation}
or equivalently as an operator on the state space:
\begin{equation}
    \hat{H}_\alpha = \alpha \sum_{n=1}^{N-1} \sum_{s=1}^d \ket{ss}_{n,n+1} \bra{ss}_{n,n+1}.
\end{equation}

A negative bias, $\alpha < 0$, encourages clustering of similar-angle plies, forming thicker blocks, while a positive bias, $\alpha > 0$, disperses plies with the same angle, making them more likely to alternate. This approach provides a probabilistic, short-range alternative to the deterministic but broader-reaching penalties we have used for the contiguity constraint, enhancing compatibility with various methods. For instance, DMRG performs more effectively with lower bond dimensions when paired with nearest-neighbor coupling terms, as do other methods like QAOA or quantum annealing, which handle lower-order interactions better.

We tested this approach with DMRG, leveraging our previous work \cite{Wulff2024} on general nearest-neighbor interactions. If the disorientation constraint is also applied, the bias $\alpha$ can be integrated directly into the matrix product operator of the disorientation constraint. \Cref{fig:progressionbias} illustrates how the stack configuration changes across a range of $\alpha$ values: for strongly negative $\alpha$, thick ply blocks dominate the stack, while for positive $\alpha$, nearly every ply has a different orientation than the previous one. This indicates that the approach is a viable and efficient method for incorporating cost-reduction considerations into the optimization.

\FloatBarrier
\clearpage
\subsection{Figures displaying the results}
\label{SuppResults}

\vspace*{\fill}
\vspace*{\fill}

\end{document}